# D4M 2.0 Schema: A General Purpose High Performance Schema for the Accumulo Database

Jeremy Kepner, Christian Anderson, William Arcand, David Bestor, Bill Bergeron, Chansup Byun, Matthew Hubbell, Peter Michaleas, Julie Mullen, David O'Gwynn, Andrew Prout, Albert Reuther, Antonio Rosa, Charles Yee

MIT Lincoln Laboratory, Lexington, MA, U.S.A.

*Abstract*— Non-traditional, relaxed consistency, triple store databases are the backbone of many web companies (e.g., Google Big Table, Amazon Dynamo, and Facebook Cassandra). The Apache Accumulo database is a high performance open source relaxed consistency database that is widely used for government applications. Obtaining the full benefits of Accumulo requires using novel schemas. The Dynamic Distributed Dimensional Data Model (D4M)[http://www.mit.edu/~kepner/D4M] provides a uniform mathematical framework based on associative arrays that encompasses both traditional (i.e., SQL) and non-traditional databases. For non-traditional databases D4M naturally leads to a general purpose schema that can be used to fully index and rapidly query every unique string in a dataset. The D4M 2.0 Schema has been applied with little or no customization to cyber, bioinformatics, scientific citation, free text, and social media data. The D4M 2.0 Schema is simple, requires minimal parsing, and achieves the highest published Accumulo ingest rates. The benefits of the D4M 2.0 Schema are *independent* of the D4M interface. *Any* interface to Accumulo can achieve these benefits by using the D4M 2.0 Schema.

*Keywords-component; D4M; NoSQL; Accumulo; database schema; Hadoop; Big Data*

I. INTRODUCTION

Non-traditional, relaxed consistency, triple store databases provide high performance on commodity computing hardware to I/O intensive data mining applications with low data modification requirements. These databases are the backbone of many web companies (e.g., Google Big Table [1], Amazon Dynamo [2,3], Facebook Cassandra [4,5], and Apache HBase [6]). The Google Big Table architecture has spawned the development of a wide variety of open source "NoSQL" database implementations [7]. Many of these implementations are built on top of the Apache Hadoop [8,9] distributed computing infrastructure that provides distributed data storage and replication services to these databases. A key element of these databases is relaxed consistency. Traditional databases provide a high level of ACID (atomicity, consistency, isolation, durability). High ACID databases guarantee that separate queries of the same data at the same time will give the same answer. Relaxed consistency databases provide BASE (Basic Availability, Soft-state, Eventual consistency), and guarantee that queries will provide the same answers eventually. In exchange, relaxed consistency databases can be built simply and provide high performance on commodity computing hardware.

The Apache Accumulo [10] database is the highest performance open source relaxed consistency database currently available and is widely used for government applications [11]. Accumulo is based on the Google Big Table architecture and formally sits on top of the Apache Hadoop distribute file system. Accumulo does not directly use the Apache Hadoop MapReduce parallel programming model. Accumulo was developed by the National Security Agency and was released to the open source community in 2011.

Obtaining the full benefits of Accumulo (and other non-traditional databases) requires using novel schemas. Traditional schema design begins with a data model and a set of target queries. The schema turns the data model into an ontology of relationships among tables with a variety of indices designed to accelerate the queries. The strengths of this approach can also cause challenges in certain applications. A data model requires a priori knowledge of the data and requires ingest processes that fully parse and normalize the data to the data model. Query optimization requires a priori knowledge of the queries so they may be captured in the table structure. Non-traditional databases allow data to be ingested and indexed with very little a priori knowledge. This allows new classes of data and queries to be added to the existing tables without modifying the schema of the database.

The Dynamic Distributed Dimensional Data Model (D4M) [12,13] provides a uniform framework based on the mathematics of associative arrays [14] that encompasses both traditional (i.e., SQL) and non-traditional databases. For non-traditional databases D4M naturally leads to a general purpose Accumulo schema that can be used to fully index and rapidly query every unique string in a dataset. The D4M 2.0 Schema builds on the D4M 1.0 schema [15] that helped inspire the widely used NuWave schema that is used across the Accumulo community. The D4M 2.0 Schema has been applied with no modification to cyber, bioinformatics, scientific citation, free text, and social media data. The D4M 2.0 Schema is simple, allows data to be ingested with minimal parsing, and the highest published Accumulo ingest rates have been achieved using this Schema [11]. The benefits of the D4M 2.0 Schema can easily be obtained using the D4M interface to Accumulo. These benefits are *independent* of the interface and *any* interface to Accumulo can achieve these benefits by using the D4M 2.0 Schema.

The organization of the rest of this paper is as follows. Section II introduces the concept of the associative array that forms the mathematical basis of the D4M 2.0 Schema. Section III presents the organization and structure of the D4M 2.0 Schema in the context of a social media example (Twitter). Section IV describes how the D4M 2.0 Schema fits into an overall data analysis pipeline.  Section V shows the performance results using Graph500 benchmark data.  Section VI summarizes the results.

## II. ASSOCIATIVE ARRAYS

Spreadsheets are used by nearly 100M people every day and may be the most commonly used analytical structure on Earth.  Likewise triple stores (e.g., Big Table, Dynamo, Cassandra, and HBase) store a large fraction of the analyzed data in the world.  Both spreadsheets and big tables can hold diverse data (e.g., strings, dates, integers, and reals) and lend themselves to diverse representations (e.g., matrices, functions, hash tables, and databases).  Despite their common usage, there have been no formal mathematics developed that can be used to describe and manipulate these data structures algebraically.

Associations between multidimensional entities (tuples) using number/string keys and number/string values can be stored in data structures called associative arrays. For example, in two dimensions, a D4M associative array entry might be

```
      A('alice ', 'bob ') = 'cited '
or    A('alice ', 'bob ') = 47.0
```

The above tuples have a 1-to-1 correspondence with their triple store representations

```
      ('alice ','bob ','cited ')
or    ('alice ','bob ',47.0)
```

Associative arrays can represent complex relationships in either a sparse matrix or a graph form (see Figure 1). Thus, associative arrays are a natural data structure for performing both matrix and graph algorithms. Such algorithms are the foundation of many complex database operations across a wide range of fields [16].

Constructing complex composable query operations can be expressed using simple array indexing of the associative array keys and values, which themselves return associative arrays:

```
   A('alice ',:)           alice row
   A('alice bob ',:)       alice and bob rows
   A('al* ',:)             rows beginning with al
   A('alice : bob ',:)     rows alice to bob
   A(1:2,:)                first two rows
   A == 47.0               subarray with values 47.0
```

The composability of associative arrays stems from the ability to define fundamental mathematical operations whose results are also associative arrays. Given two associative arrays `A` and `B`, the results of all the following operations will also be associative arrays

```
   A + B    A - B    A & B    A|B    A*B
```

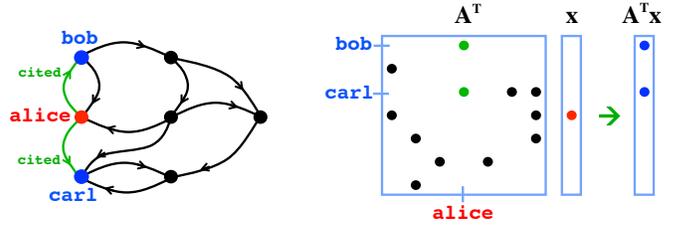

Figure 1. A graph describing the relationship between `alice`, `bob`, and `carl` (left). A sparse associative array **A** captures the same relationships (right). The fundamental operation of graphs is finding neighbors from a vertex (breadth first search). The fundamental operation of linear algebra is vector matrix multiply. D4M associative arrays make these two operations identical.  Thus, algorithm developers can simultaneously use both graph theory and linear algebra to exploit complex data.

Associative array composability can be further grounded in the mathematical closure of semirings (i.e., linear algebraic "like" operations) on multidimensional functions of infinite, strict, totally ordered sets (i.e., sorted strings).  Associative arrays when combined with fuzzy algebra [17,18,19] allows linear algebra to be extended beyond real numbers to include words and strings.  For example, in standard linear algebra, multiplying the vector $\mathbf{x}$ = (`'alice bob '`) by the vector $\mathbf{y}$ = (`'carl bob '`) is undefined.  In fuzzy algebra we can replace the traditional plus (+) operation with a function like "max" and the traditional multiply operation with an operation like "min" resulting in

$$\mathbf{x}\,\mathbf{y}^\mathrm{T} = (\text{'alice bob '})(\text{'carl bob '})^\mathrm{T}$$
$$= \max(\min(\text{'alice carl '}),\min(\text{'bob bob '}))$$
$$= \max(\text{'alice bob '})$$
$$= \text{'bob '}$$

where $^\mathrm{T}$ denotes the transpose of the vector.  Using fuzzy algebra allows D4M to apply much of the extensive mathematics of linear algebra to an entirely new domain of data consisting of words and strings (e.g., documents, network logs, social media, and DNA sequences). Measurements using D4M indicate these algorithms can be implemented with a tenfold decrease in coding effort when compared to standard approaches [20,21].

## III. D4M 2.0 SCHEMA

The D4M 2.0 Schema is best explained in the context of a specific example. Twitter is a micro-blog that allows its users to globally post 140 character entries or "tweets." Twitter has over 100M users who produce 500M tweets per day.  Each tweet consists of a message payload and metadata.  To facilitate social media research the NIST Tweets2011 corpus [22,23] was assembled consisting of 16M tweets over a two-week period in early 2011.  At the time of our harvesting this corpus it consisted 161M distinct data entries from 5.3M unique users.  The entire Tweets2011 corpus was ingested into the D4M 2.0 schema running on a single node Accumulo instance in about 20 minutes, corresponding to an ingest rate of >200K entries/second.

**Accumulo Tables:**

Tedge/TedgeT

| Row Key | stat\|200 | stat\|301 | stat\|302 | stat\|403 | ... | time\|2011- | time\|2011- | time\|2011- | time\|null | user\|bimo... | user\|Mich... | user\|Pen... | user\|... | ... | word\|@mi... | word\|null | word\|Tipo. | word\|Você | word\|Wait |
|---|---|---|---|---|---|---|---|---|---|---|---|---|---|---|---|---|---|---|---|
| 08805831972220092 | ■ | | | | | ■ | | | ■ | ■ | | | | | ■ | ■ | | | |
| 75683042703220092 | ■ | | | | | | ■ | | | ■ | | | | | | | | | ■ |
| 08822929613220092 | | ■ | | | | | | ■ | | | | | | | | | ■ | | |
| ... | | | | | | | | | | | | | | | | | | | |

TedgeDeg^T

Row Key

| Degree | 108 642 73 | 286 150 7 | 836 327 | 825 822 | ... | 6 | 7 | 7 | 454 596 8 | 6 | 7 | 3 | | ... | 3 | 454 603 9 | 16 | 102 23 | 162 4 |

TedgeTxt

| Row Key | text |
|---|---|
| 08805831972220092 | @mi_pegadejeito Tipo. Você fazer uma plaquinha pra mim, com o nome do FC pra você tirar uma foto, pode fazer isso? |
| 75683042703220092 | Wait :) |
| 08822929613220092 | null |
| ... | |

Figure 2. The D4M 2.0 Schema as it is applied to Twitter data consists of four tables. The raw tweet text is stored in one column in the TedgeTxt table. All the meta data (stat|, user|, time|) and the parsed text (word|) are stored in Tedge such that each column|value pair is a unique column. Storing the transpose of the metadata in TedgeT creates an index to every unique string in the dataset and allows it to be looked up in a few milliseconds. The sums of the unique column|value pairs are stored using an accumulator column labeled Degree in the TedgeDeg table. The sum table enables efficient query planning by allowing queries to estimate the size of their results prior to executing queries. The row keys are flipped to allow for efficient load balancing as the table grows and is split across multiple servers.

The simplest way to view Accumulo is as a triple store of strings consisting of a row key, a column key, and a value that correspond to the entries of a sparse matrix. In Accumulo terminology these are the row, column qualifier, and value (Accumulo has additional row properties that will be discussed shortly). In the case of twitter, a triple might be

(31963172416000001,user|getuki,1)

The above triple denotes that the flipped tweet id 31963172416000001 was from the user getuki. As is often the case in the D4M 2.0 Schema the value of 1 is used to simply denote the existence of the relationship and the value itself has no additional meaning.

Figure 2 shows the D4M 2.0 Schema applied to the Tweets2011 data resulting in four distinct tables. The raw tweet text is stored in one column in the TedgeTxt table. All the meta data (stat|, user|, time|) and the parsed text (word|) are stored in Tedge such that each column|value pair is a unique column. Storing the transpose of the metadata in TedgeT indices every unique string the dataset allows it to be looked up in a few milliseconds. The sums of the unique column|value pairs are stored using an accumulator column labeled Degree in the TedgeDeg table. The sum table enables efficient query planning by allowing queries to estimate the size of results prior to executing queries. The row keys are stored in flipped format to allow for efficient load balancing as the table grows and is split (or sharded) across multiple servers.

The specific features of Accumulo and how they are exploited by the D4M 2.0 Schema are as follows.

*A. Row Store*

Accumulo is a row store so any row key (e.g., the tweet ID 31963172416000001) can be looked up in constant time. However, looking up a column (e.g., user|getuki) or value (e.g., 1) requires a complete scan of the table. The D4M 2.0 Schema addresses this limitation by storing both the table (Tedge) and its transpose (TedgeT), allowing any row or column to be looked up in constant time.

*B. Sparse*

Accumulo storage is sparse. Only non-empty columns are stored in a row. This is critical since many of the data sets that Accumulo are used on are naturally represented as extremely sparse tables. In the Tweet2011 data 99.99997% of the 16M x 30M sparse matrix is empty.

*C. Unlimited Columns*

Accumulo can add new columns with no penalty. This is a key capability of Accumulo that is heavily exploited by the D4M 2.0 Schema. It is often the case that there will be more unique columns than rows. Tweet2011 has 16M unique rows and 30M unique columns.

*D. Arbitrary Text*

Accumulo rows, columns, and values can be arbitrary byte strings. This is very useful for storing numeric data (e.g., counts) or multi-lingual data (e.g., unicode). For example, consider the following Twitter entry

```
TweetID          stat time                user  text
10000061427136913 200  2011-01-31 06:33:08 getuki バスなう
```

In a traditional database, the above entry would be represented by one row in a four column table. In the Accumulo D4M 2.0

Scheme this entry is represented in the `Tedge` table by the following four triples with flipped row keys

    (31963172416000001,stat|200,1)
    (31963172416000001,time|2011-01-31 06:33:08,1)
    (31963172416000001,user|getuki,1)
    (31963172416000001,word|バスなう,1)

Note: since this tweet has only one word the `text` field is parsed into just one `word`. The raw entry for this tweet would likewise be stored in the `TedgeTxt` table as

    (31963172416000001,text,バスなう)

This raw table allows all data to be preserved in case the original context of the tweet is desired (as is often the case).

*E. Collective Updates*

Accumulo performs collective updates to tables called "mutations" that can update many triples at the same time. It is often optimal to have thousands of triples in a single mutation. In the Tweets2011 data, inserts were performed in batches of 10,000 tweets to achieve optimal performance.

*F. Accumulators*

Accumulo can modify values at insert time. For example, if the following triple were inserted into the `TedgeDeg` table

    (word|バスなう,Degree,1)

and the table entry already had a value of

    (word|バスなう,Degree,16)

then Accumulo can be instructed that any such collision on the column `Degree` should be handled by converting the strings `16` and `1` to numeric values, adding them, and then converting them back to a string to be stored as

    (word|バスなう,Degree,17)

An accumulator column is used to create the `TedgeDeg` column sum table in the D4M 2.0 Schema. The `TedgeDeg` sum table provides several benefits. First, the sum table allows tally queries like "how many tweets have a specific word" to be answered trivially. Second, the sum tables provides effective query planning. For example, to find all tweets containing two words, one first queries to the sum table to select the word that is the least popular before proceeding to query the transpose table (`TedgeT`).

**Note**: Directly inserting all triples into the sum table can create a bottleneck. During large ingests into Accumulo it is vital to pre-sum the columns in each batch prior to ingesting into the sum table. Pre-summing can reduce the traffic into the sum table by 10x or more. In the D4M API pre-summing can be achieved by constructing an associative array `A` of all the triples in the batch and then simply inserting the result of `sum(A,2)` into `TedgeDeg`.

*G. Parallel*

Accumulo is highly parallel. At any given time it is possible to have many processes inserting and querying the database. Even on a single node, the optimal ingest performance for the Tweets2011 data was achieved using 4 ingest processes on the same node.

*H. Distributed*

Accumulo uses distributed storage. As Accumulo tables become large they are broken up into pieces called tablets that can be stored on different tablets.

*I. Partitions*

Accumulo tables are partitioned (or sharded) into different tablets at specific row keys that are called splits. As a table increases in size Accumulo will automatically pick splits that keep the pieces approximately equal. If the row key has a time like element to it (as does the tweet ID), then it is important to convert it to a flipped format so the most rapidly changing digits are first. This will cause inserts to be spread across all the tablets. If the row key is sequential in time, all the inserts will be performed on only one tablet and then slowly migrated to the other tablets. Avoiding this "burning candle" effect is critical to achieve high performance in Accumulo.

Because the size at which Accumulo starts splitting tables is quite large, it is often necessary to pre-split the table to achieve optimal performance. Pre-splitting is even important on single node databases (see Figure 5).

*J. Hadoop without MapReduce*

Accumulo lives on top of the Apache Hadoop Distributed File System (HDFS) [9]. Apache Hadoop also provides an interface to the simple and very popular MapReduce parallel programming model. Unfortunately, many applications require more sophisticated programming models to achieve high performance [24]. Accumulo does not use the Hadoop MapReduce parallel Java API for any of its internal operations. The highest performance Accumulo applications typically do not use Hadoop MapReduce to execute their ingest and query programs. In this paper the results were obtained using either the pMatlab [25,26,27] parallel programming environment that uses a distributed arrays parallel programming model or the LLGrid_MapReduce interface that leverages the GridEngine parallel computing scheduler [11].

*K. Accumulo Advanced Features*

Accumulo has a number of additional features. For most applications, viewing Accumulo entries as triples is sufficient. In reality, an Accumulo entry is not a 3-tuple (row, column, value) but a 6-tuple (row, column family, column qualifier, visibility label, timestamp, value). The D4M 2.0 Schema chooses to view Accumulo entries as triples because it simplifies table design, takes full advantage of Accumulo's core features (high performance and unlimited columns), and is compatible with the 2D tabular view found in nearly every database. The D4M 2.0 Schema views the other parts of the Accumulo entry (column family, visibility label, timestamp) as additional metadata on the triples that can used when needed.

Every Accumulo entry is coded with a timestamp that allows an entry to hold its entire history of values. The most common use of the timestamp is in Accumulo's internal automated data role-off that cleanses the system of older values. Role-off is the most common way data is truly deleted from the Accumulo system. Until role-off, data is typically only marked for deletion and still remains on the system. The specific role-off policy is application dependent and is typically set by the database administrator. Actual time data is best held in a column (e.g., `time|2011-01-31 06:33:08`). A time column can be queried and manipulated like any other data.

Accumulo visibility labels provide a sophisticated Boolean algebra for each entry to determine who can see each piece of data. In some systems visibility labels are essential, in other systems visibility labels are unused. Visibility policy is usually application dependent and is typically set by the database administrator.

Accumulo column families provide an additional hierarchy to the Accumulo columns for accelerating certain operations. Columns families can be kept together on the same table split, which can increase the performance of certain queries that always have data together. In addition, Accumulo iterators have been written that allow joining data more efficiently. The D4M 2.0 Schema provides these benefits through the use of the sum table `TedgeDeg`. It is tempting to use the Accumulo column family to convey one additional level of semantic hierarchy, but this can have unintended performance implications. The D4M 2.0 Schema embeds arbitrary levels of semantic of hierarchy directly into the column (e.g., `stat|200` and `user|getuki`). This column format has the added advantage of being easily represented as a row in the transpose table `TedgeT`.

A final Accumulo advanced feature is bulk ingest. Bulk ingest stages data in the internal Accumulo format and adds it to the database as part of its internal bookkeeping processes. In certain instances bulk ingest may provide higher performance at the cost of delaying when the data is available for query. The highest published ingest rates [11] use Accumulo's standard ingest mechanism with batched mutations that make the data immediately available for query.

## IV. PIPELINE

Accumulo databases do not run in isolation and are usually a part of a data analysis pipeline. The D4M 2.0 Schema typically has a four-step pipeline consisting of parse, ingest, query/scan, and analyze. The parse step converts the raw data (e.g., CSV, TSV, or JSON format) to simple triples. In addition, each batch of triples is also saved as a D4M associative array.

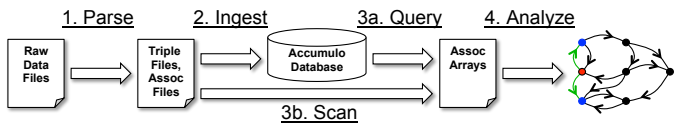

Figure 3. D4M 2.0 Schema pipeline consists of parse, ingest, query/scan, and analyze steps.

The ingest step reads the triple files and ingests them into the `Tedge`, `TedgeT`, and `TedgeTxt` tables. In addition, the associative array files are read in, summed and the results added to the sum table `TedgeDeg`. Extraction of data for analysis is done either by querying the data directly or by scanning the associative array files. If the amount of data required for the analysis is small then querying the database will be fastest. If the amount of data required is a large fraction of the entire database (>10%) then it is often faster to run the analysis in parallel over the associative array files.

## V. PERFORMANCE RESUTLS

The ingest performance of the D4M 2.0 Schema on an 8-node (192 core) system is taken from [11] and shown in Figure 4. The best published performance results we were able to find for Cassandra [28] and HBase [29] are also shown. Figure 4 is consistent with the claim that Accumulo is the highest performance and most scalable open source triple store database currently available. Likewise, Figure 4 is also consistent with the claim that the D4M 2.0 Schema is currently the highest performance Accumulo schema.

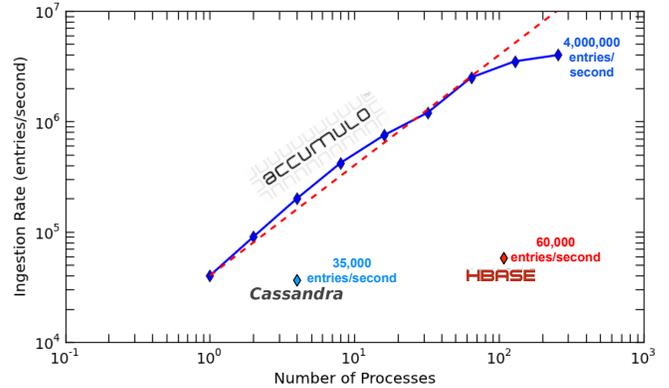

Figure 4. Ingest performance vs. number of ingest processors of Accumulo using the D4M 2.0 Schema [11], Cassandra [28], and HBase [29].

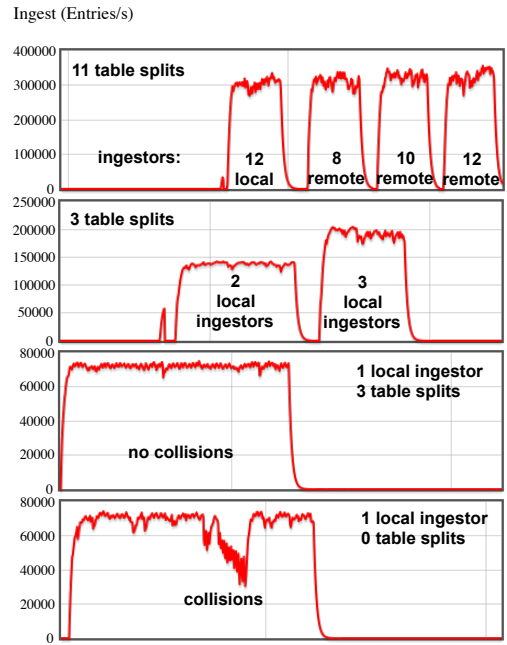

Figure 5. D4M 2.0 Schema performance on a 1-node (32 core) Accumulo system ingesting Graph500 data [31,32]. Using pre-splitting a single ingestor is able to achieve sustained performance of 70K inserts/sec. Likewise using multiple ingestors and pre-splits it is possible achieve nearly 350K inserts/sec.

Figure 5 is a more detailed analysis of the performance of the D4M 2.0 Schema on a 1-node (32 core) system using Graph500 data [30,31]. Figure 5 (bottom) is the ingest performance of a single ingestor with no pre-splitting. A drop occurs halfway thru the ingest due to collisions. Even though only one ingestor is used, a very high performance D4M ingestor can post mutations to Accumulo faster than it can retire them. Figure 5 (bottom middle) shows that by adding splits a sustained performance of 70K (entries/sec) can be achieved by a single ingestor. Figure 5 (top middle) shows the sustained performance for 2 ingestors (140K entries/sec) and 3 ingestors (190K entries/sec). Figure 5 (top) shows the single node performance peaking a ~350K entries/sec)

## VI. Summary

The D4M 2.0 Schema is simple and requires minimal parsing. The highest published Accumulo ingest rates have been achieved using this schema. The benefits of the schema have been illustrated using the 16M record Tweets2011 corpus which was fully parsed, ingested, and indexed in <1 hour with <1 day of programmer effort. Additional performance results are shown using the Graph500 benchmark that achieved an insert rate ~350K entries/sec on a single node database while preserving constant (subsecond) access time to any entry. The benefits of the D4M 2.0 Schema are independent of the D4M interface and any interface to Accumulo can achieve these benefits by using the D4M 2.0 Schema.